# Molecularly Resolved Electronic Landscapes of Differing Acceptor-Donor Interface Geometries

*Katherine A. Cochrane,[1,3] Tanya S. Roussy,[2] Bingkai Yuan,[3] Gary Tom,[2,3] Erik Mårsell,[3,4] and Sarah A. Burke\*[1,2,3]*

[1] Department of Chemistry, University of British Columbia, Vancouver British Columbia, Canada, V6T 1Z1

[2] Department of Physics and Astronomy, University of British Columbia, Vancouver British Columbia, Canada, V6T 1Z1

[3] Stewart Blusson Quantum Matter Institute, University of British Columbia, Vancouver British Columbia, Canada, V6T 1Z4

[4] Division of Molecular and Condensed Matter Physics, Department of Physics and Astronomy, Uppsala University, Box 516, 751 20 Uppsala, Sweden

\* corresponding author: saburke@phas.ubc.ca

ABSTRACT:

Organic semiconductors are a promising class of materials for numerous electronic and optoelectronic applications, including solar cells. However, these materials tend to be extremely sensitive to the local environment and surrounding molecular geometry, causing the energy levels near boundaries and interfaces essential to device function to differ from those of the bulk. Scanning Tunneling Microscopy and Spectroscopy (STM/STS) has the ability to examine both the structural and electronic properties of these interfaces on the molecular and submolecular scale. Here we investigate the prototypical acceptor/donor system PTCDA/CuPc using sub-molecularly resolved pixel-by-pixel STS to demonstrate the importance of subtle changes in interface geometry in prototypical solar cell materials. PTCDA and CuPc were sequentially deposited on NaCl bilayers to create lateral heterojunctions that were decoupled from the underlying substrate. Donor and acceptor states were observed to shift in opposite directions suggesting an equilibrium charge transfer between the two. Narrowing of the gap energy compared to isolated molecules on the same surface are indicative of the influence of the local dielectric environment. Further, we find that the electronic state energies of both acceptor and donor are strongly dependent on the ratio and positioning of both molecules in larger clusters. This molecular-scale structural dependence of the electronic states of both interfacial acceptor and donor has significant implications for device design where level alignment strongly correlates to device performance.

INTRODUCTION:

Organic photovoltaic (OPV) materials offer great promise for low-cost, lightweight devices with low embodied energy. However, even with recent advances leading to photo-conversion efficiency surpassing 14%, OPV devices have not made sustained headway into the mass energy market.[1] The current dominant architecture for both vacuum[2,3] and solution[4] processed devices relies on the engineered energy landscape at interfaces between acceptor and donor materials to drive separation of the strongly bound exciton into free carriers.[5,6] As a result, understanding this interface and how energy level alignment depends on the structure of the interface and surrounding molecular material is crucial for understanding and improving device performance.[7] As the soft inter- and intra-molecular interactions permit a wide range of possible structures and conformations,[8,9] these interfaces have the potential to be highly inhomogeneous on the molecular scale where exciton dissociation takes place. Meanwhile, different interfacial structures have the potential to lead to different energy level alignments. [8,10-13] For example, Graham, et al. demonstrated that device performance could be significantly altered with subtle changes in interfacial molecular geometry driven by altering functional groups. [13]

Many techniques that allow for simultaneous electronic and spatial resolution are diffraction limited, and cannot probe structure and properties on the length scale on which the processes for charge transfer and separation occur. These key processes rely significantly on device structure on the molecular scale.[14] Scanning probe microscopy (SPM) offers the ability to resolve molecular and sub-molecular structure as well as the local density of states yielding energy level alignment on Ångström scales.[15-17] Previous SPM studies on closed monolayer systems of mixed donor-acceptor prototypical organic semiconductors have shown shifts in the energy levels of both components of the mixed domains.[12,18-21] These shifts were attributed to a combination of interactions, including non-covalent intermolecular interactions, as well as intermolecular and molecule-substrate charge transfer. In these monolayer studies, the geometry of the interface was shown to be important, however the systems were formed by self-assembly which limits the variety of available structures. In addition, these systems were often directly adsorbed on metallic substrates where strong interactions with the underlying surface complicate understanding of the intermolecular interactions that matter for the organic–organic interfaces in OPVs.

Here, we use pixel-by-pixel scanning tunneling microscopy and spectroscopy (STM/STS) as well as non-contact atomic force microscopy (NC-AFM) to probe small clusters of acceptor and donor molecules to investigate how energy level alignment is influenced by variations in molecular-scale interface geometry. The experiment was performed on a thin salt film on Ag(111) to prevent hybridization with the underlying metal substrate. Taking advantage of small island sizes, in conjunction with molecular manipulation, we are able to probe specific geometries that cannot be isolated within closed monolayer systems. Probing a variety of structures shows changes in energy levels that occur with only subtle differences in interface geometry. Pthalocyanines and perylene derivatives serve as well-studied model acceptor–donor combinations,[12,22-25] tracing back to the original bilayer heterojunction solar cell reported by Tang, et al.[2] However, the relation between interfacial energy alignment and precisely controlled interfacial geometry has yet to be determined. Understanding this relationship between molecular and electronic structure is essential for maximizing the efficiency of charge transfer and separation, which in turn facilitates the design of new and improved materials and devices.

METHODS

Small clusters of 3,4,9,10-peryelene tetracarboxylic dianhydride (PTCDA) and Copper(II) Phthalocyanine (CuPc) on NaCl(2ML)/Ag(111) were prepared in ultrahigh vacuum (UHV). Thin salt layers were used to reduce the influence of the underlying metal substrate.[26] All STM and STS measurements were taken at ~4.3 K and $<5 \times 10^{-12}$ mbar with a low temperature scanning probe microscope (LT-SPM, Scienta Omicron) using a cut platinum-iridium tip. Differential conductance data was acquired by numerically differentiating I(V) curves obtained at every (x,y) position with the feedback loop disabled during the sweep. All spectra were processed to minimize the influence of the transmission function as well as the divergence near the Fermi energy resulting from normalization (see supporting information for details).[27,28] NC-AFM measurements were taken by measuring the frequency shift in constant height mode with a QPlus sensor with a resonance frequency of 27.3 kHz and a Q of ~18,000. NaCl (TraceSELECT ≥ 99.999%, Fluka) bilayers were formed by thermal deposition at ~530 °C with the Ag(111) (Mateck GmbH) substrate held at approximately 80 °C resulting in (001) terminated islands with about 50% coverage. PTCDA (98%, Alfa Aesar) and CuPc (>99.95%, Aldrich) were sequentially thermally deposited at 325 °C and 400 °C respectively after *in situ* outassing, with the sample held at ~4.4 K. The sample was annealed slightly by removing it from the cryostat (in UHV) for several minutes, allowing the sample to warm and form a variety of small aggregates (Fig. 1a). CuPc molecules were occasionally manipulated with the tip to form interfaces,[29] otherwise self-assembled formations were examined.

RESULTS AND DISCUSSION

Figure 1a shows an STM topographic overview of PTCDA and CuPc molecules sequentially deposited on NaCl(2ML)/Ag(111) and subsequently annealed at room temperature for two minutes. After this anneal, small aggregates of 2 to ~6 molecules were observed, as well as isolated PTCDA. Isolated CuPc molecules on NaCl were observed after the initial deposition but not after annealing, indicating that they diffuse when the substrate is warmed. At biases less than -0.4 V and greater than 1 V, PTCDA is imaged with STM as a double lobed structure on NaCl(2ML)/Ag(111) (Figure 1b), while the in gap states of PTCDA are imaged as a single lobe between -0.4 V and 1 V (Figure 1c) as found previously.[10] PTCDA adsorbs on the $Cl^-$ top site on NaCl(2ML)/Ag(111).[30][31] CuPc is imaged with STM as a variety of 16, 8, and 4 fold symmetric structures at different biases, with the in gap states appearing as a cross. High resolution NC-AFM indicates that CuPc adsorbs on or near the $Na^+$ top-site, similar to CuPc on NaCl bilayers on Cu(100).[32]

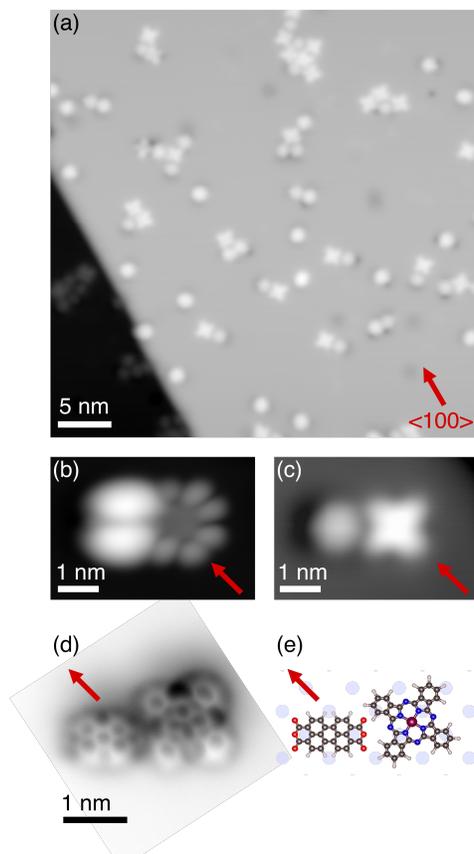

**Figure 1.** (a) STM topographic image showing typical cluster sizes and geometry of PTCDA and CuPc on NaCl(2ML)/Ag(111) NaCl <100> vectors as determined from the NaCl island edges are represented by red arrows (40 nm x 40 nm, $I_t$ = 5 pA, $V_b$ = +0.5 V). (b,c) STM topography of a PTCDA (left) and CuPc (right) heterodimer (5 nm x 3.5 nm, $I_t$ = 6 pA, b. $V_b$ = -2 V and c. $V_b$ = +0.5 V). (d) High resolution NC-AFM frequency shift image of a PTCDA/ CuPc heterodimer acquired with a PTCDA functionalized tip (3.5 nm x 3.5 nm, 0 V, constant z). (e) PTCDA and CuPc positioning roughly determined from the NC-AFM image and proposed adsorption geometry on the underlying NaCl lattice.

One of the most common structures observed was the heterodimer seen in Figures 1 b – e. STM imaging (Figures 1b, c) shows the gross intermolecular arrangement, while high resolution NC-AFM (Figure 1d) reveals a slight rotation of the CuPc with respect to the NaCl lattice. Due to this rotation, two hydrogens of one of the CuPc phenyl rings are positioned adjacent to one of the PTCDA anhydride oxygens. This results in the PTCDA molecule being closer to one of the CuPc phenyl rings. Although NC-AFM gives an indication of intra- and intermolecular distances, precise determinations are influenced by distortions caused by the interaction of the PTCDA functionalized tip with the molecules on the surface as well as deformations of the adsorbed molecule due to interactions with the substrate.[33] Nevertheless, the distances between the CuPc hydrogens and the PTCDA oxygens suggest that weak hydrogen bonding between the two molecules is likely. The PTCDA molecule can be switched between two equivalent NaCl lattice sites with respect to the CuPc molecule by applying a voltage pulse with the tip over the molecule.

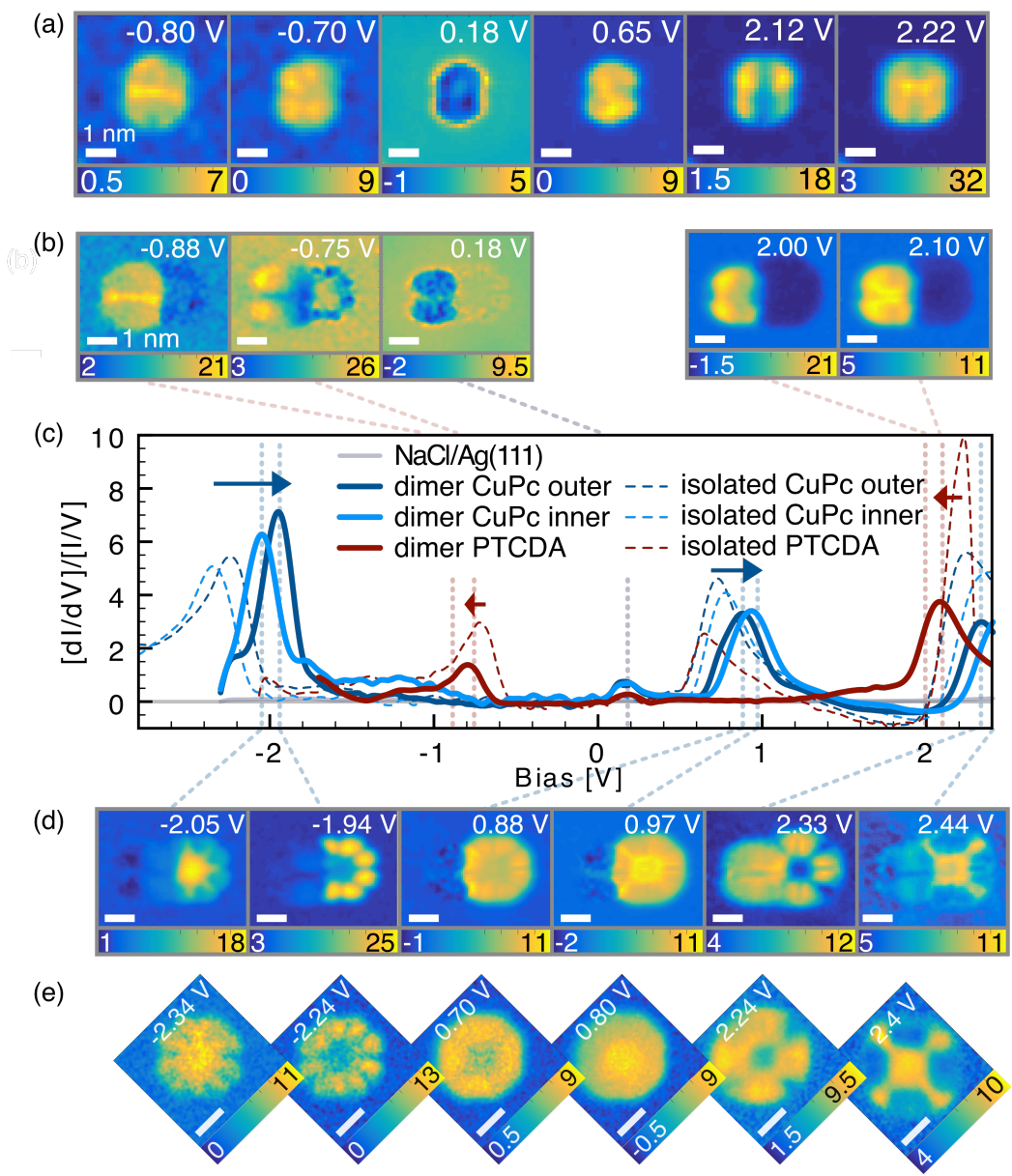

**Figure 2.** (a) STS maps of an isolated PTCDA on NaCl(2ML)/Ag(111) at increasing sample bias (5 nm x 5 nm, set-point parameters $I_t$= 2.5 pA, $V_b$= -2 V). (b) STS maps of a PTCDA/CuPc heterodimer corresponding to the PTCDA states at increasing sample bias, indicated with vertical red dashed lines in (c), (5 nm x 4 nm, set-point $I_t$= 2 pA, $V_b$= -2.5 V). (c) [dI/dV]/[I/V] point spectra of both isolated CuPc and PTCDA molecules (dashed lines) as well as PTCDA and CuPc in the heterodimer (solid lines). Arrows indicate direction of shifting of energy levels upon heterodimer formation. (d) STS maps of a PTCDA/CuPc heterodimer corresponding to CuPc states at increasing sample bias, indicated with vertical blue dashed lines in (c), (5 nm x 4 nm). (e) STS maps of an isolated CuPc on NaCl(2ML)/Ag(111) at increasing sample bias (3.5 nm x 3.5 nm, setpoint $I_t$= 2 pA, $V_b$= -3.1 V). The images are rotated to match the orientation of the CuPc molecule in (d).

To understand the influence of different heterojunction geometries, STS measurements of clusters were compared to the isolated molecules. STS of isolated PTCDA (Figure 2a and 2c, dashed red line) show three strong resonances, which we label P-O1, P-U1 and P-U2: P-O1 at -0.70 V with a shoulder at -0.80 V, P-U1 at +0.65 V, and P-U2 with a shoulder at 2.12 V and a peak at 2.22 V. PTCDA on NaCl(2ML)/Ag(111) is expected to be negatively charged due to the small work function of NaCl(2ML)/Ag(111) and comparatively large electron affinity of PTCDA.[10] This results in the splitting of the singly occupied LUMO into the upper and lower Hubbard states, lying above and below the Fermi energy ($E_F$).[34,35] From the STS point spectra and the spatial maps (Figure 2a and c), we assign P-O1 as the overlapping HOMO and LUMO$^{-1\rightarrow 0}$ (the lower Hubbard state), P-U1 as the LUMO$^{1-\rightarrow 2-}$ (the upper Hubbard state), and P-U2 as the nearly degenerate LUMO+1/LUMO+2.[10] The states seen at +0.18 V around the negatively charged PTCDA molecule in the isolated molecule as well as the coordinated system appear due to the scattering of the NaCl(2ML)/Ag(111) surface state, which appears as a step at ~0.1 V.[36]

For an isolated CuPc molecule, STS measurements (Figure 2c dashed blue lines and 2e) show occupied and unoccupied resonances alternating between the metal center (labeled "inner") and the outer phthalocyanine macrocycle (labeled "outer"). The peaks and shoulders of the point STS as well as the corresponding spatial distributions in the maps allow for determination of several unique molecular states, which we label C–O1 and C–O2 for the occupied states, and C–U1 to C–U4 for the unoccupied states. Directly assigning STS resonances on metal phthalocyanines on insulating surfaces to specific molecular orbitals has proven to be nontrivial, requiring formalisms that consider many-body interactions.[37] In CuPc this is partially influenced by the singly occupied 4s orbital resulting in spin splitting.[38] Several theoretical studies of CuPc have been performed, however orbital sequence changes significantly with the type of theory used.[38-42] Results from even the most computationally intensive and detailed theory published for CuPc[38] do not satisfactorily compare to our data in terms of the ordering of states and their corresponding symmetries. Because of this, we do not identify these resonances by specific molecular orbitals, but rather refer to the observed resonances in terms of these empirically assigned occupied and unoccupied levels in the order they appear in STS.

Tunneling spectra and maps of the PTCDA/CuPc heterodimer are also shown in Figure 2 and compared to the maps of the isolated PTCDA and CuPc molecules shown in Figures 2a and e respectively. The STS maps of each molecule in the heterodimer are spatially similar to corresponding resonances in the isolated molecules. This similar orbital appearance and symmetry allows identification of the corresponding states between the heterodimer and each isolated molecule, and indicates a relatively weak interaction between the two molecules. In addition to the similarity of the spatial maps, there is no noticeable broadening in the dominant PTCDA peaks P-O1 and U2 and all of the CuPc states. This is indicative of minimal electronic hybridization between the molecules in the heterodimer. There is a noticeable broadening, reduction in intensity, and inward shift of the P-U1 state (upper Hubbard state), sometimes leading to difficulty in identifying the energy of this state. The simplest interpretation of the broadening would be that it arises from electronic delocalization within the molecular cluster. However, the states of the adjacent CuPc remain unaffected, and the spatial distribution seen in STS maps likewise remains distinct, leading us to conclude minimal electronic hybridization between the two molecules. We attribute the behavior of the PTCDA U-1 to the on-site electrostatic repulsion (Hubbard potential) due to screening by neighboring molecules, however the intensity, shape and width of this resonance in STS data under different conditions has yet to be explained.

The changes in energy level alignment between isolated molecules and those in the heterodimer system are compared in more detail in Figure 3. The STS spectra of a PTCDA molecule adjacent to a CuPc show an overall downward shift in energy as compared with an isolated PTCDA. Likewise, there is an overall upward shift of the CuPc spectra in the heterodimer relative to an isolated molecule. Both molecules also show a shift of the states towards $E_F$. The negative (downward) shift of PTCDA is indicative of a loss of negative charge. Both the ionization potential (IP) and the electron affinity (EA) of PTCDA in the heterodimer have increased. The positive (upward) shift of CuPc indicates the addition of negative charge, with a smaller IP and EA. Although PTCDA is typically thought of as an electron acceptor, here the initial negative charge on the PTCDA molecule results in electron donation to the CuPc. Due to the lack of intermolecular hybridization, we do not believe that the shifted energy levels come from a transfer of a full unit of charge, but a time average exchange predominantly due to electron hopping, either directly between molecules or mediated by the underlying metallic substrate. This observed partial charge transfer can be attributed to the timescale of the STS measurement. which averages over the fluctuating occupation of each molecule.

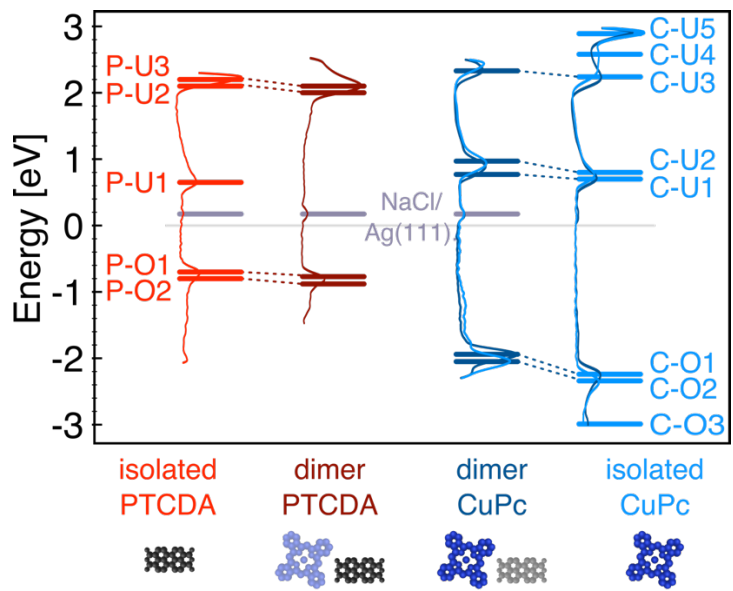

Figure 3. Electronic states of isolated PTCDA (red), isolated CuPc (cyan), PTCDA in a heterodimer (maroon), and CuPc in a heterodimer (blue) determined from STS (also shown). State positions are indicated by thick bars, and shifts are indicated with dashed lines. Position of the NaCl (2ML)/ Ag(111) surface state is indicated by the gray bar.

The inward shift of the observed states towards $E_F$ are consistent with effects due to the local polarization environment.[43,44] As the observed shifts occur in the heterodimer relative to the isolated molecule on the same substrate, the changes must be related to interactions between the two molecules. Previously we have shown that in-plane intermolecular polarization energies can shift energy levels inwards by 100's of meV.[10] This is due to charges added or removed from one molecule in tunneling or in transport being stabilized by an induced dipole in the other. The energy shift due to polarization was determined from the measured gap sizes ($\Delta$) by $E_P \approx (\Delta_{isolated} - \Delta_{heterodimer})/2$, assuming $E_P$ is equal for the different states probed. The experimentally determined induced energy shift of PTCDA due to the adjacent CuPc in the heterodimer, $E_P$, was found to be ~35 meV and the energy shift of CuPc due to PTCDA was ~70 meV.

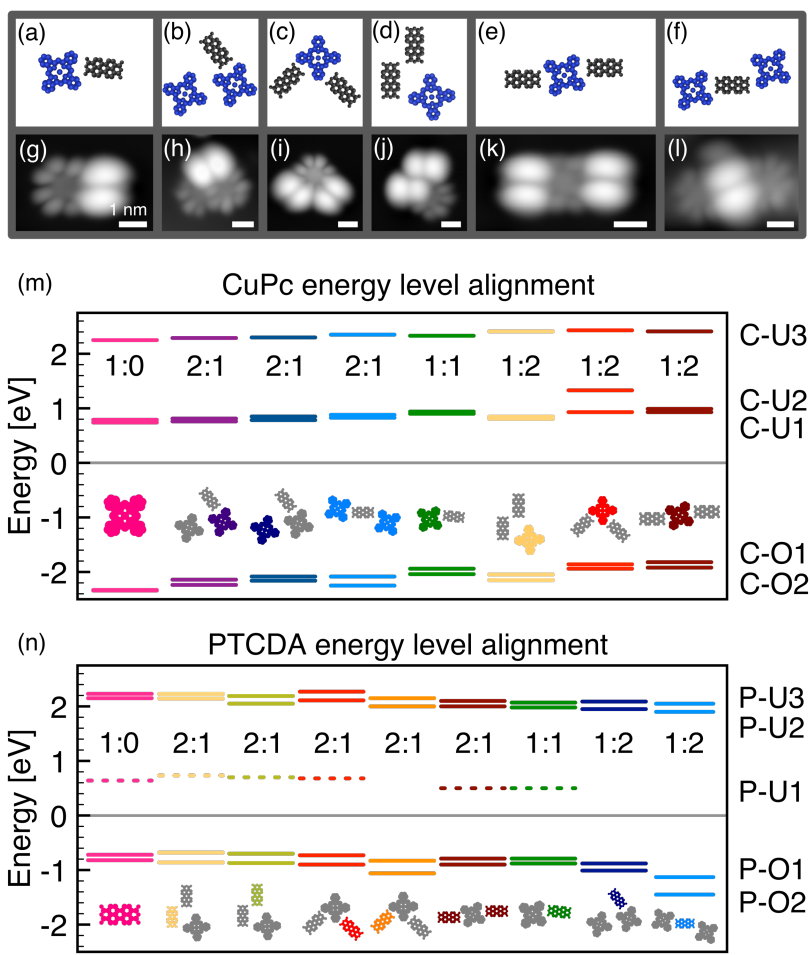

Figure 4. Comparison of energy level alignment of PTCDA/CuPc clusters of varying geometries. (a-f) Structural models of the 6 clusters examined. (g-l) STM topographic images of each cluster $I_t$ = 2 pA (g: 4 x 5 nm $V_B$=-2V, h: 6 x 6 nm -2.1V, i: 5.5 x 5.5 nm -2.2 V, j: 5 x 5 nm -2.1 V, k: 6.5 x 3.5 nm -2V, and l: 5.5 x 4 nm -2.1 V. (m) Energy alignment of CuPc states arranged in increasing order of surrounding PTCDA molecules. (n) Energy alignment of PTCDA states arranged in increasing order of surrounding CuPc molecules. In (m) and (n) the stoichiometry of the cluster is indicated with respect to the molecule examined.

In order to investigate the influence of the interface structure on energy level alignment, we performed STS measurements on six different geometries of PTCDA/CuPc interfaces (Figure 4a – l). Each molecule was in a different local environment within clusters with varying acceptor–donor stoichiometry and geometrical arrangement. The energy levels for each molecule were identified from the STS measurements and are shown for each CuPc and PTCDA in Figure 4m–n, which reveals the gross features present. For both molecules, the displayed energy levels are ordered by the number of surrounding acceptor or donor molecules. As a PTCDA molecule is surrounded by more CuPc molecules, the energy levels generally shifted down, indicating more negative charge is transferred from the PTCDA (Figure 4n). For the same stoichiometry, the energy levels are also generally shifted down where the PTCDA-CuPc distances are smaller. For CuPc, an increasing number of adjacent PTCDA molecules generally results in an increasing upward shift in energy, indicating a gain in negative charge (Figure 4m). In some cases, it was not possible to accurately assign an energy value to the PTCDA upper Hubbard state (P–U1) due to significant reduction in intensity and possible broadening, as discussed above and previously noted in PTCDA clusters.[10]

Islands with equivalent numbers of each molecule but with differing molecular geometries still show substantial differences, indicating that the precise arrangement of molecules matters as well as the local stoichiometry at the interface. For example, in the two PTCDA:CuPc 1:2 clusters with different structures (Figure 4b/h and f/l), the PTCDA O1 state is significantly shifted down (Figure 4n) in the linear island (f/l) compared with the elbow-shaped (b/h) island. This results in PTCDA O1-U2 gap energies that are ~200 meV different for the two different orientations. In the PTCDA:CuPc 2:1 clusters, the PTCDA U1 state is significantly shifted and broadened in some geometries (shown in more detail in Figure 5c). This stoichiometry also results in the largest variation in CuPc gap energies of ~150 meV depending on the geometry.

Even more striking, islands containing the same heterodimer unit (i.e. with the same intermolecular distance and relative orientation), still show differences in energy level alignment due to the differing surrounding environment. For example, the clusters with CuPc:PTCDA stoichiometry of 2:1 (Figure 4b/g and 4e/h) both have this heterodimer unit with the only difference being additional surrounding CuPc molecules. The CuPc molecules that are the same distance from a negative PTCDA molecule but with different neighbors show energy level shifts that differ by ~100 meV. In these cases, the overall Coulombic field the molecules experience is the same for all three configurations due to the presence of only one negatively charged PTCDA molecule. This indicates that the field generated by the presence of a negative point charge cannot fully account for this energy shift, and the surrounding molecular geometry of neutral species must be taken into account.

To provide a more detailed visualization of the energy landscapes of the molecular clusters Figure 5 shows the STS profiles along the indicated paths across each of the molecular islands examined in Figure 4, with distance along path on the x-axis, bias on the y-axis and the color-scale representing (dI/dV)/(I/V) intensity. We observe that the electronic resonances are strongly localized over the corresponding molecule however, a slight overlap in the interfacial region between the acceptor and donor molecules is present. Similar overlap was also observed in a covalently bonded bipolar intermolecular heterojunction both in the experimental data and the corresponding DFT. In that case, the width of the overlap was determined to be roughly twice the length of the covalent bond between the acceptor and donor moieties. The PTCDA-CuPc heterojunctions we observe have a similar degree of overlap (~3-6 Å) as measured by STS although there is no clear evidence or expectation of a covalent bond in this case.

In addition, it can be seen in Figure 5 that the energy levels over an individual molecule are not perfectly flat, i.e. the energy landscape for charge addition and removal change on sub-molecular length scales. There is a consistent slight bending away from the Fermi energy on the CuPc resonances moving closer to adjacent PTCDA molecules. The addition of a nearby point charge would result in a bending of the vacuum level and thereby an overall upward or downward shift in energy of the occupied and unoccupied resonances. Due to the asymmetry of the bending observed in the hetero-clusters, we do not attribute this shift to a Coulombic field due to the initial charge on the PTCDA, resulting in a symmetric bend in the vacuum level. Likewise, near-field polarization effects would be expected to stabilize charge and bend these levels towards $E_F$, indicating that these subtle shifts are not likely electrostatic in origin. These fine near-field effects are unlikely to influence charge separation and transport as the grosser features above, though they do indicate that neighboring molecular geometry influences the energy level landscape on even sub-molecular length scales.

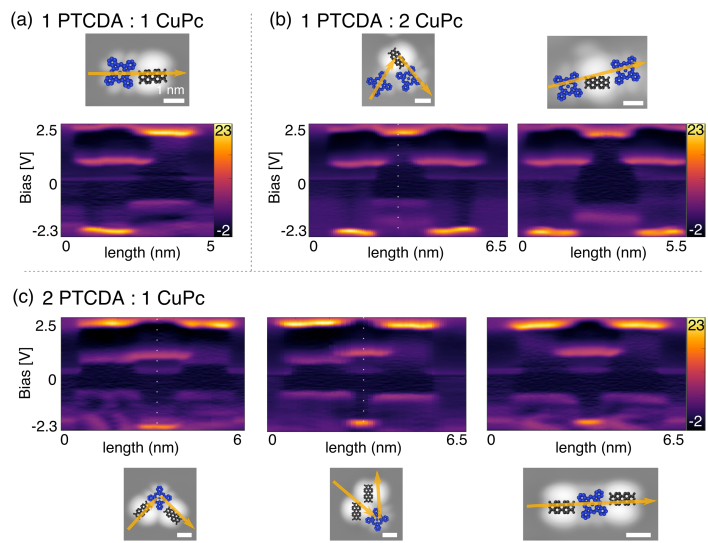

Figure 5: STS along a line cut (indicated by yellow arrows) of bimolecular heterojunctions of islands consisting of varying eometries of (a) one CuPc and one PTCDA, (b) two CuPc and one PTCDA (c) one CuPc and two PTCDA. x-axis is distance along line cut, y-axis is bias, and colormap is (dI/dV)/(I/V) intensity. White dotted line indicates node of line path where applicable.

CONCLUSION

In summary we show that significant changes in electronic structure, and thus energy level alignment, can occur at acceptor–donor interfaces due to local stoichiometry and geometry on a single molecule level. The effect of stoichiometry is striking in cases where one of the components is charged, as PTCDA is here. The small differences in interfacial molecular geometry studied in varying heteromolecular islands of PTCDA and CuPc show further significant effects on the electronic structure, including changes in gap energy of up to 200 meV. Because the energy level alignment at the heteromolecular interface is expected to be the driving force for exciton dissociation, it is crucial to gain a molecular scale understanding of how these subtle geometrical changes alter the energetics of these interfaces. Recently, Henneke et. al, have demonstrated that geometric control over the interface is possible for the PTCDA/CuPc system. Different ratios of the initial components result in different geometries in self-assembled monolayers.[22] By changing deposition parameters, a specific structure can be chosen. Combining these and other self-assembly strategies, including addition of functional groups as part of the full molecular design, with the knowledge gained from probing energy level alignment, one has the potential to optimize interfacial morphology to create new device structures with significant improvements in performance.

## ASSOCIATED CONTENT

## SUPPORTING INFORMATION

Supporting information is available that further describes the STS normalization and processing.

## NOTES

The authors declare no competing financial interests.

## ABBREVIATIONS

2 ML, two monolayer; 2D, two dimensional; CuPc, copper phthalocynanine; EA, activation energy; $E_F$, Fermi energy; $E$p, polarization energy; H-bond, hydrogen bond; HOMO, highest unoccupied molecular orbital; IP, ionization potential; LT, low temperature; LUMO, lowest unoccupied molecular orbital, NC-AFM, non-contact atomic force microscopy; PTCDA, 3,4,9,10-perylenetetracarboxylic dianhydride; SPM, scanning probe microscopy; STM, scanning tunneling microscopy; STS, scanning tunneling spectroscopy; UHV, ultra-high vacuum.

## ACKNOWLEDGEMENTS:

The authors gratefully acknowledge support from the NSERC Discovery grants program No. 402072-2012, the Canadian Foundation for Innovation, the NSERC CREATE-QUEST grant No. 414110-12 (T.R. and S.B.), the Canada Research Chairs Program (S.B.) This work was also supported by the Swedish Research Council (VR) International Postdoc Grant 2016-06719 (E.M.).

TOC GRAPHIC

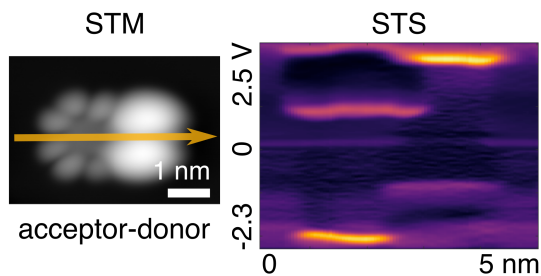